\documentclass[lettersize,journal]{IEEEtran}
\usepackage{cite}
\usepackage{enumerate}
\usepackage{amsmath}
\usepackage{amsfonts}
\usepackage{mathrsfs}
\usepackage{amssymb}
\usepackage{amsmath,amsfonts}
\usepackage{algorithmic}
\usepackage{algorithm}
\usepackage{array}
\usepackage[caption=false,font=normalsize,labelfont=sf,textfont=sf]{subfig}
\usepackage{color}
\usepackage{textcomp}
\usepackage{stfloats}
\usepackage{url}
\usepackage{verbatim}
\usepackage{graphicx}
\usepackage{cite}
\usepackage[mathscr]{euscript}
\usepackage{booktabs}
\usepackage{bbm}
\usepackage{amssymb}
\usepackage{glossaries}
\usepackage{bm}
\usepackage{balance}

\usepackage{setspace} 
\newacronym{AN}{AN}{artificial noise}
\newacronym{AWGN}{AWGN}{additive white Gaussian noise}
\newacronym{BS}{BS}{base station}
\newacronym{CRB}{CRB}{Cram\'{e}r-Rao bound}
\newacronym{CSI}{CSI}{channel state information}
\newacronym{CU}{CU}{communication user}
\newacronym{DFRC}{DFRC}{dual-functional radar-communication}
\newacronym{Eve}{Eve}{eavesdropper}
\newacronym{fFAMA}{\textit{f}-FAMA}{fast fluid antenna multiple access}
\newacronym{FAS}{FAS}{fluid antenna system}
\newacronym{sFAMA}{\textit{s}-FAMA}{slow fluid antenna multiple access}
\newacronym{ISAC}{ISAC}{integrated sensing and communications}
\newacronym{LoS}{LoS}{line-of-sight}
\newacronym{MI}{MI}{mutual information}
\newacronym{MIMO}{MIMO}{multiple-input multiple-output}
\newacronym{MISO}{MISO}{multiple-input single-output}
\newacronym{SCA}{SCA}{successive convex approximation}
\newacronym{SDR}{SDR}{semidefinite relaxation}
\newacronym{SIR}{SIR}{signal-to-interference ratio}
\newacronym{SNR}{SNR}{signal-to-noise ratio}
\newacronym{SINR}{SINR}{signal-to-interference-plus-noise ratio}
\newacronym{ULA}{ULA}{uniform linear array}
\newacronym{DoF}{DoF}{degrees of freedom}

\begin{document}

%\title{Joint Power and Motion Control for mmWave and Sub-6GHz Dual-Mode UAV Communications}
\title{Shifting the ISAC Trade-Off with Fluid Antenna Systems}
\author{Jiaqi Zou,~\IEEEmembership{Graduate Student Member,~IEEE}, Hao Xu,~\IEEEmembership{Member,~IEEE,} Chao Wang,~\IEEEmembership{Senior Member,~IEEE}, \\Lvxin Xu, Songlin Sun,~\IEEEmembership{Senior Member,~IEEE}, Kaitao Meng,~\IEEEmembership{Member,~IEEE}, Christos Masouros,~\IEEEmembership{Fellow,~IEEE}, and Kai-Kit Wong,~\IEEEmembership{Fellow,~IEEE}
        % <-this % stops a space
\thanks{Jiaqi Zou, Lvxin Xu, and Songlin Sun are with the School of Information and Communication Engineering, Beijing University of Posts and Telecommunications (BUPT), Beijing 100876, China, and also with the Key Laboratory of Trustworthy Distributed Computing and Service (BUPT), Ministry of Education, China, and also with the National Engineering Laboratory for Mobile Network Security, BUPT, Beijing, China. (E-mail:\{jqzou, xulvxin, slsun\}@bupt.edu.cn). }
\thanks{Hao Xu, Kaitao Meng, Christos Masouros, and Kai-Kit Wong are with the Department of Electronic and Electrical Engineering, University College London, London WC1E 6BT, United Kingdom. (E-mail: hao.xu, kaitao.meng, c.masouros, kai-kit.wong@ucl.ac.uk).}
\thanks{Chao Wang is with Integrated Service Networks Lab, Xidian University, Xi'an 710071, China. (E-mail: drchaowang@126.com).} 
%\thanks{Kaitao Meng and Christos Masouros are with the Department of Electronic and Electrical Engineering, University College London, London, UK (emails: {kaitao.meng, c.masouros}@ucl.ac.uk).}
}

% The letter headers
\markboth{}%
{Shell \MakeLowercase{\textit{et al.}}: A Sample Article Using IEEEtran.cls for IEEE Journals}


\newtheorem{Th}{Theorem}
\newtheorem{Lm}{Lemma}
\newtheorem{propRemark}{Remark}
\maketitle

\begin{abstract}
As an emerging antenna technology,  a fluid antenna system (FAS) enhances spatial diversity to improve both sensing and communication performance by shifting the active antennas among available ports. In this letter, we study the potential of shifting the integrated sensing and communication (ISAC) trade-off with FAS. We propose the model for FAS-enabled ISAC and jointly optimize the transmit beamforming and port selection of FAS.  In particular, we aim to minimize the transmit power, while satisfying both communication and sensing requirements. An efficient iterative algorithm based on sparse optimization, convex approximation, and a penalty approach is developed.
The simulation results show that the proposed scheme can attain 33\% reductions in transmit power with guaranteed sensing and communication performance, showing the great potential of the fluid antenna for striking a flexible tradeoff between sensing and communication in ISAC systems.
\end{abstract}

\begin{IEEEkeywords}
Fluid antenna system, integrated sensing and communications, port selection.
\end{IEEEkeywords}

\section{Introduction}

As one of the emerging technologies in next-generation wireless networks, \gls{ISAC} offers a sustainable way of supporting these two previously separate functionalities \cite{liu2022integrated}. Various design methodologies have been proposed in recent years, aiming to unlock the potential of  \gls{ISAC} with shared signal processing techniques and even hardware devices. 
ISAC system designers aim to meet both the specific sensing and communications (S\&C) requirements and balance the conflicting S\&C objectives. For instance, the fundamental tradeoffs between the communication sumrate and \gls{CRB} for target parameter estimation were studied in \cite{liu2021cram}.  Besides, a novel performance metric was proposed in \cite{zou2023energy} to measure the sensing-centric energy efficiency, and the trade-off between communication-centric and sensing-centric energy efficiency was also investigated.

The unveiled potential of ISAC is also attributed to the development of \gls{MIMO} and massive MIMO due to its multiplexing and diversity gain. Currently, as an emerging antenna technology, \gls{FAS} exploits more favorable channel conditions and spatial diversity incorporating fluidity in the antenna technology through shifting the active antenna \cite{FluidAntenna_Survey,xu2023channel}. This introduces additional degrees of freedom to improve communication performance.  
Despite the growing attention for FAS in the area of antenna design, the potential of FAS in promoting wireless communication has not been explored until the work of \cite{Wong_Fluid}.
% the fluid antenna has received increasing attention in the antenna community over the past decade, it was not until the work of   that the potential of applying the fluid antenna to improve wireless communication performance was investigated.
More recently, \cite{wong2021FAMA} utilized a switchable antenna port and proposed a novel \gls{fFAMA} with the spatial diversity offered by FAS, which requires fast port switch on a symbol-by-symbol basis.
% to exploit the sum-interference signal null. 
To reduce the complexity of \gls{fFAMA}, \gls{sFAMA} was proposed in \cite{wong2023slow} which only needs to switch the port of the fluid antenna when the channel changes. In \cite{xu2023outage,xu2024revisiting}, a two-user FAMA system was considered and the outage performance was investigated under a fully correlated channel model.
To further reduce the system overhead consumed by \gls{CSI} estimation of all the ports, \cite{chai2022port} studied a deep-learning-based algorithm. Furthermore, \cite{zou2023online} proposed a bandit-learning-based framework for online port selection, eliminating the requirements of instantaneous full CSI.

Although FAS has demonstrated significant improvements in communication-only systems, its ability to support ISAC, especially for the sensing functionality, has not been well investigated. Compared with the conventional fixed antenna selection for ISAC\cite{valiulahi2022antenna}, FAS allows for shifting more active ports within a constrained space, providing unique flexibility of reconfigurable radiating elements.
The deployment of FAS facilitates the advanced management of propagation environments, thereby yielding increased degrees of freedom (DoF) for enhancing the integration gain in ISAC systems, e.g., choosing an antenna position with higher channel correlation between sensing and communication~\cite{meng2024integrated}.
In this letter, we investigate FAS-enabled ISAC, where we consider a FAS \gls{BS} that simultaneously achieves multicast communication and target sensing. In particular, our objective is to minimize the transmit power, considering the \gls{SNR} requirement for sensing,  the \gls{SNR} requirement for communication, and the constraint of available ports. Our aim is to explore the enhancements in the ISAC trade-off that the FAS capability offers. Whereas the considered problem is NP-hard, we propose an iterative optimization algorithm leveraging sparse optimization and linear matrix inequality to transform the non-convex problem into a sequence of convex optimization problems. Simulation results show that the proposed method for the FAS-assisted ISAC system leads to a significant reduction in power consumption and demonstrates the great potential advantage for improving the flexibility of balancing the S\&C performance ISAC.
%to fulfill the increasing demand for communication.

\section{System Model}
\begin{figure}
	\centering
	\includegraphics[width=0.8\linewidth]{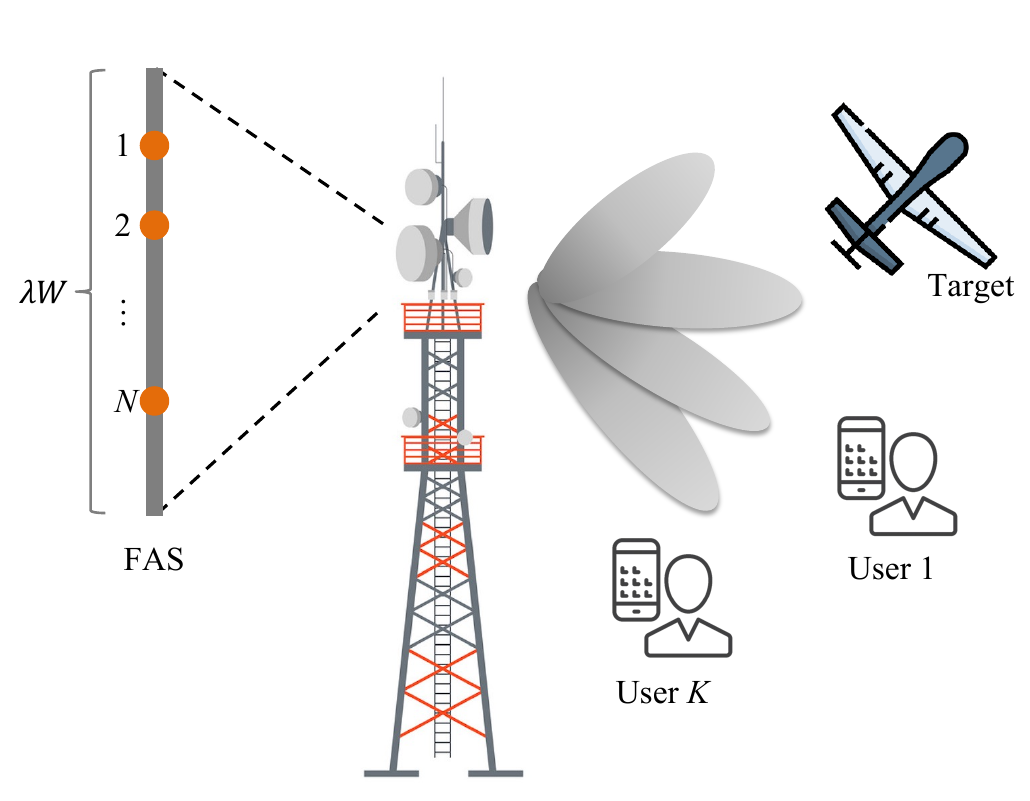}
	\caption{System model.}
	\label{fig:systemmodel}
\end{figure}

\subsection{System Setting}

As shown in Fig.~\ref{fig:systemmodel},  we consider a system with a multicast ISAC BS and $K$ mobile users. The BS transmits a common signal to the users and utilizes the echo signals to perform potential target sensing. Each mobile user has a fixed-position antenna. Differently, the BS is equipped with an $N$-antenna FAS, where $M$ predetermined ports are uniformly distributed along a linear space of length $\lambda W$. Here, $\lambda$ is the signal wavelength and $W$ is the normalized size of FAS. $N$ different antennas can change their positions among the $M$ available ports such that the ISAC performance can be enhanced. 
Compared with the traditional \gls{MIMO},  the port switching capability of \gls{FAS} gives more spatial DoF to the \gls{ISAC} \gls{BS}.
\subsection{Multicast Communication Performance}
Let $s \sim {\cal {CN}} (0,1)$ denote the communication signal to be transmitted. The received signal at the $k$th  user is given by
\begin{align}
	y_k=\mathbf{h}_k^H\mathbf{w}s +n_k,
\end{align}
where $\mathbf{h}_k \in \mathbb{C}^{M\times 1}$, $\mathbf{w}\in\mathbb{C}^{M\times 1}$, and $n_k\sim\mathcal{CN}\left(0,\sigma_c^2\right)$ denote the downlink communication channel from the \gls{BS} to the $k$th user, the \gls{ISAC} beamformer, and the \gls{AWGN} received at the $k$th user, respectively.
We also assume that $N$ antennas are available, i.e., only $N$ of $M$ ports are activated to generate the ISAC waveform. Thus, $\mathbf{w}$ satisfies $||\mathbf{w}||_0=N$, where $\left\| \cdot\right\| _0$  represents the $L_0$ norm.
For the communication channel, we follow \cite{new2023information} to express the spatial correlation among the ports as $\mathbf{J}$. Decomposing $\mathbf{J}$, we have $\mathbf{J} =\mathbf{U}_{\mathrm{tx}} \bm{\Lambda}_{\mathrm{tx}}^H \mathbf{U}_{\mathrm{tx}}^H$, where $\bm{\Lambda}_{\mathrm{tx}}\in\mathbb{C}^{M\times M}$ is a diagonal matrix whose elements are the eigenvalues of spatial correlation matrix $\mathbf{J}$ and the columns of $\mathbf{U}_{\mathrm{tx}}\in\mathbb{C}^{M\times M}$ are the corresponding eigenvectors. Then, the spatially correlated channel can be given as
\begin{align}
	\mathbf{h}_k = \mathbf{g}^H\sqrt{\bm{\Lambda}_{\mathrm{tx}}^H}\mathbf{U}_{\mathrm{tx}}^H,
\end{align} 
%where $\bm{\Lambda}_{\mathrm{tx}}\in\mathbb{C}^{M\times M}$ is a diagonal matrix whose elements are the eigenvalues of spatial correlation matrix $\mathbf{J}$ and the columns of $\mathbf{U}_{\mathrm{tx}}\in\mathbb{C}^{M\times M}$ are the corresponding eigenvectors.
where $\mathbf{g} \sim \mathcal{CN}\left(\mathbf{0},\mathbf{I}_M\right)$ reprensents the random slow fading coefficient.

Accordingly, the communication \gls{SNR} at the $k$th user is given by
\begin{align}
	\gamma_{c,k}= \frac{\left|\mathbf{h}_k^H\mathbf{w}\right|^2}{\sigma_c^2}.
\end{align}
\subsection{Target Sensing Performance}
For sensing, we consider a colocated monostatic
MIMO radar system equipped with $N_r$ receive antennas.
Then, the echo signal received at the sensing receiver is given by
\begin{align}
	\mathbf{y}_r=\mathbf{A}(v)\mathbf{w}s+\mathbf{z}_r,
\end{align}
where $\mathbf{A}(v)=\beta\bm{\alpha}_r(v_r)\bm{\alpha}_t^T(v_t)$ and $\mathbf{z}_r\sim\mathcal{CN}\left(\mathbf{0},\sigma_r^2\mathbf{I}_{N_r}\right)$ denote the target response matrix and the \gls{AWGN}, respectively. $\beta$ represents the reflection coefficient.
%, which contains both the round-trip path-loss and the radar cross section of the target.
$\bm{\alpha}_t(v_t)=\left[1,\ldots,e^{-j2\pi\left(N_t-1\right)\Delta_t\mathrm{sin}(v_r)}\right]$ and
$\bm{\alpha}_r(v_r)=\left[1,\ldots,e^{-j2\pi\left(N_t-1\right)\Delta_r\mathrm{sin}(v_r)}\right]$ denote the transmitting and receiving steering vectors, respectively, where $\Delta_t$ and $\Delta_r$ are the antenna spacing of transmit and receive antennas, respectively. Here, we consider the straightforward scenario of single-target sensing for the initial exploration of FAS-ISAC. However, our proposed method can be extended to multiple-target sensing by incorporating reflections from multiple targets. As we consider the monostatic colocated MIMO radar, we have the identical angle of departure (AOD) and angle of arrival (AOA) of the target following the existing literature~\cite{zou2023energy,liu2021cram}, denoted as $v_t = v_r = v$.

It is worth noting that we adopt the fixed antenna array at the sensing receiver for angle estimation\footnote{The self interference can be suppressed by hardware design, e.g., antenna separation, and self interference cancellation algorithm\cite{7105651}.}. Accordingly, the average sensing \gls{SNR}  is formulated as
\begin{align}
	\gamma_{r}=\frac{||\mathbf{A}(v)\mathbf{w}||_2^2}{\sigma_r^2}.
\end{align}

\section{Joint Port Selection and Dual-Functional Beamforming Optimization}
Based on the above models, we now turn to designing a joint port selection and beamforming scheme for FAS-enabled ISAC. Our objective is to minimize the total transmit power under the constraints of the minimum sensing \gls{SNR} requirement $r_s$, minimum communication \gls{SNR} requirement $r_c$, and the number of activated ports. The considered problem can be formulated as 
\begin{subequations}\label{JointPortSelectionBeamforming}
	\begin{align}
		&\underset{\mathbf{w}}{\mathrm{minimize}}\; ||\mathbf{w}||_2^2\\
		&\mathrm{s.t.}\; \gamma_{r}\geq r_s,\; \gamma_{c,k}\geq r_c, k\in\mathcal{K},\; ||\mathbf{w}||_0=N, \label{SNR_Constraint}
	\end{align}
\end{subequations}
where $\mathcal{K}=\left\{1,\ldots, K\right\}$ denotes the communication user set.
Different from traditional multicast \gls{ISAC} systems, the additional sparse constraint enforces that only $N < M$ antenna ports are activated to generate \gls{ISAC} signal.

Problem \eqref{JointPortSelectionBeamforming} is challenging to handle, since the constraints in (\ref{SNR_Constraint}) are all non-convex. Besides,  the port selection problem is a nonconvex combinatorial optimization problem due to sparse constraint in (\ref{SNR_Constraint}), which makes problem (\ref{JointPortSelectionBeamforming}) more challenging to handle. 
To deal with the problem (\ref{JointPortSelectionBeamforming}), we adopt sparse optimization to propose an iterative optimization algorithm \cite{MulticastISAC}.
Firstly, we adopt a penalty method to move non-convex sparse constraint into the objective function and obtain the following penalized problem
\begin{subequations}\label{nonconvexproblem}
\begin{align}
	&\underset{\mathbf{w}}{\mathrm{minimize}}\;||\mathbf{w}||_2^2+\lambda ||\mathbf{w}||_0\\
	&\mathrm{s.t.}\;  \gamma_{r}\geq r_s,\; \gamma_{c,k}\geq r_c,\;k\in\mathcal{K},\label{nonconvexconstraintJoint}
\end{align} 
\end{subequations}
where $\lambda$ is a penalized parameter that can balance between the objective of minimizing the number of the selected ports and that of minimizing the total power. Therefore, we can adjust $\lambda$ to select $N$ ports. However, $||\mathbf{w}||_0$ is still non-convex, which makes the objective function difficult to handle. As an alternative, considering that the $L_1$ norm, $||\mathbf{w}||_1$, serves as a convex approximation of $||\mathbf{w}||_0$, we resort to the convex approximation of problem (\ref{nonconvexproblem}) for obtaining its local optimal solution, i.e.,
\begin{align}
	&\underset{\mathbf{w}}{\mathrm{minimize}}\;||\mathbf{w}||_2^2+\lambda ||\mathbf{w}||_1 \;\;\mathrm{s.t.}\; (\ref{nonconvexconstraintJoint}).\label{convexapproximation1}
\end{align}
For each $\lambda$, we can find $\hat{\lambda}$ to reformulate problem (\ref{convexapproximation1}) as 
\begin{align}
	&\underset{\mathbf{w}}{\mathrm{minimize}}\;||\mathbf{w}||_2^2+\hat{\lambda} ||\mathbf{w}||_1^2 \;\;\mathrm{s.t.}\; (\ref{nonconvexconstraintJoint}).\label{convexapproximation}
\end{align}
Since the objective function of problem (\ref{convexapproximation}) is a convex approximation of that of problem (\ref{nonconvexproblem}), we can solve problem (\ref{convexapproximation}) iteratively to obtain a local optimal solution of problem (\ref{nonconvexproblem}) \cite{ConvexApproximation}. However, the constraints in problem (\ref{convexapproximation}) are still nonconvex, making the problem nontractable.
To reformulate problem (\ref{convexapproximation}) as a tractable one, we introduce an auxiliary matrix $\mathbf{X}=\mathbf{w}\mathbf{w}^H$ to reformulate problem (\ref{convexapproximation}) as 
\begin{subequations}\label{Rankconstraintproblem}
	\begin{align}
		&\underset{\mathbf{X}}{\mathrm{minimize}}\;\mathrm{Tr}\left(\mathbf{X}\right)+\hat{\lambda}\mathrm{Tr}\left(\mathbf{1}_{N\times N}\left|\mathbf{X}\right|\right)\\
		&\mathrm{s.t.}\;  \frac{\mathrm{Tr}\left( \mathbf{h}_k^H\mathbf{X}\mathbf{h}_k\right) }{\sigma_c^2}\geq r_c,\; \frac{\mathrm{Tr}\left(\mathbf{A}(v)\mathbf{X}\mathbf{A}^H(v)\right)}{\sigma_r^2}\geq r_s,\label{new_constraint2}
		\\
		&\qquad\mathrm{rank}\left(\mathbf{X}\right)=1, \label{rank_constraint}
	\end{align} 
\end{subequations}
where the reformulation of the objective function is due to the fact that $||\mathbf{w}||_1^2=\mathbf{1}^T_{N\times 1}|\mathbf{X}|\mathbf{1}_{N\times 1}=\mathrm{Tr}\left(\mathbf{1}_{N\times N}\left|\mathbf{X}\right|\right)$.
We observe that problem (\ref{Rankconstraintproblem}) is still non-convex because of the rank-1 constraint (\ref{rank_constraint}). To tackle this problem,  we exploit  \cite[Lemma 1]{zou2023energy} and reformulate problem (\ref{Rankconstraintproblem}) as
\begin{subequations}
	\begin{align}
		&\underset{\mathbf{X},\mathbf{w}}{\mathrm{minimize}}\;\mathrm{Tr}\left(\mathbf{X}\right)+\hat{\lambda}\mathrm{Tr}\left(\mathbf{1}_{N\times N}\left|\mathbf{X}\right|\right)\\
		&\mathrm{s.t.}\;  \left[\begin{matrix}
			\mathbf{X},&\mathbf{w}\\
			\mathbf{w}^H,&1
		\end{matrix}\right]\succeq \mathbf{0},\;\mathrm{Tr}\left(\mathbf{X}\right)\leq \mathbf{w}^H\mathbf{w},\\
	&\qquad (\ref{new_constraint2}).
	\label{Reformulated_Problem}
	\end{align} 
\end{subequations}
Although problem (\ref{Reformulated_Problem}) is still non-convex, we can further adopt \gls{SCA} to approximate it as a sequence of convex problems, which are given by
\begin{subequations}\label{Reformulated_Problem2}
\begin{align}
		&\underset{\mathbf{X},\mathbf{w}}{\mathrm{minimize}}\;\mathrm{Tr}\left(\mathbf{X}\right)+\hat{\lambda}\mathrm{Tr}\left(\mathbf{1}_{N\times N}\left|\mathbf{X}\right|\right)\\
&\mathrm{s.t.}\;  \left[\begin{matrix}
	\mathbf{X},&\mathbf{w}\\
	\mathbf{w}^H,&1
\end{matrix}\right]\succeq \mathbf{0},\;\mathrm{Tr}\left(\mathbf{X}\right)\!\leq \!-\mathbf{w}^H_{l-1}\mathbf{w}_{l-1}\!+\!2\mathrm{Re}\left(\mathbf{w}^H_{l-1}\mathbf{w}\right),\label{newnewconstraint}\\
&\qquad (\ref{new_constraint2}).
\end{align}
\end{subequations}
Therefore, we can adopt bisection search over $\hat{\lambda}$ to solve a sequence of problems (\ref{Reformulated_Problem2}) for getting a local optimal solution to problem (\ref{JointPortSelectionBeamforming}). To further increase the sparsity of $\mathbf{w}$, we adopt the iteratively re-weighted $l_1$-norm penalty in \cite{Sparisity}. 
For completeness, we sketch the iteratively re-weighted $l_1$-norm penalty approach. 
In particular, by introducing an auxiliary matrix $\mathbf{U}$, we  construct a new penalty problem as follows
\begin{subequations}\label{Reformulated_Problem2_1}
	\begin{align}
		&\underset{\mathbf{X},\mathbf{w}}{\mathrm{minimize}}\;\mathrm{Tr}\left(\mathbf{X}\right)+\hat{\lambda}\mathrm{Tr}\left(\mathbf{U}\left|\mathbf{X}\right|\right)\\
		&\mathrm{s.t.}\;  (\ref{newnewconstraint}),\; (\ref{new_constraint2}).\label{newconstraint3}
	\end{align}
\end{subequations}

\begin{algorithm}[!t]
	\algsetup{linenosize=\footnotesize}
	%\scriptsize
	\caption{{Iteratively Re-Weighted $l_1$-Norm Penalty Approach.}} % 
	\label{JointOptimziationAlgorithm1}
	\begin{algorithmic}[1]
		\STATE Set $j=0$ and initialize $\mathbf{U}^{(j)}=\mathbf{1}_{N\times N}$;
		\WHILE{1}
		\WHILE {not converged}
		\STATE Solve the following weighted $l_1$-norm problem to obtain $\mathbf{X}^*$ and $\mathbf{w}^*$
		\begin{align}
	&\underset{\mathbf{X},\mathbf{w}}{\mathrm{minimize}}\;\mathrm{Tr}\left(\mathbf{X}\right)+\hat{\lambda}\mathrm{Tr}\left(\mathbf{U}^{(j)}\left|\mathbf{X}\right|\right),\;\mathrm{s.t.}\;(\ref{newconstraint3}) \notag
		\end{align}
		\STATE Update $\mathbf{w}_{l-1}=\mathbf{w}^*$
	\ENDWHILE
	\STATE Output $\mathbf{X}^*$
	\STATE Update  the weight matrix $\mathbf{U}^{j+1}(m,n)=1/\left(\left|\mathbf{X}^{*}(m,n)\right|+\epsilon\right)$ for each $m,n=1,\ldots,M$	
		\STATE $j=j+1$, if $j$ exceeds the threshold, output $\mathbf{U}^*$, break.
	\ENDWHILE		 				
	\end{algorithmic}
\end{algorithm}
\noindent The iterative algorithm is given in Algorithm 1. After obtaining $\mathbf{U}^*$, we adopt bisection search algorithm to find $\lambda$ for making $||\mathbf{w}||_0=N$, which is given by Algorithm 2.
\begin{algorithm}[!t]
	\algsetup{linenosize=\footnotesize}
	%\scriptsize
	\caption{{Proposed Joint Port Selection and \gls{ISAC} Beamforming Optimization.}} % 
	\label{JointOptimziationAlgorithm2}
	\begin{algorithmic}[1]
		\STATE Initialize $\hat{\lambda}_L$ and $\hat{\lambda}_U$. Set the tolerance $\epsilon\ll 1$ and $\hat{\lambda}=\frac{\lambda_U+\lambda_L}{2}$
		\WHILE{$\hat{\lambda}_U-\hat{\lambda}_L>\epsilon$}
         \WHILE {not converged}
         \STATE Solve the following weighted $l_1$-norm problem to obtain $\mathbf{X}^*$ and $\mathbf{w}^*$
         \begin{align}
         	&\underset{\mathbf{X},\mathbf{w}}{\mathrm{minimize}}\;\mathrm{Tr}\left(\mathbf{X}\right)+\hat{\lambda}\mathrm{Tr}\left(\mathbf{U}^{(*)}\left|\mathbf{X}\right|\right),\;\mathrm{s.t.}\;(\ref{newconstraint3}) \notag
         \end{align}
         \STATE Update $\mathbf{w}_{l-1}=\mathbf{w}^*$
         \ENDWHILE
         \STATE If $||\mathbf{w}||_0>N$, set $\hat{\lambda}_L=\hat{\lambda}$, else set $\hat{\lambda}_U=\hat{\lambda}$. 
		\ENDWHILE		 				
	\end{algorithmic}
\end{algorithm}
\section{Simulation Results}

In this section, we provide the simulation results of the proposed joint beamforming and port selection algorithm for FAS-enabled ISAC.
Unless stated otherwise, the available ports on the fluid antenna are set to $M = 32$ with $W$ = 2. Compared to conventional uniformly-distributed antenna systems, which are limited to only four antennas with half-wavelength spacing within the restricted space of $2\lambda$, FAS permits the implementation of a larger number of active antennas and more flexible position among the ports. Besides, we assume that $\sigma^2_c = \sigma^2_r = 1$, and $K = 10$. For sensing, we consider a point-like target located at $v_t = v_r = v = 60^\circ$ with the reflection coefficient $\beta = 0.1$. 
%度

\begin{figure}
	\centering
	\includegraphics[width=0.95\linewidth]{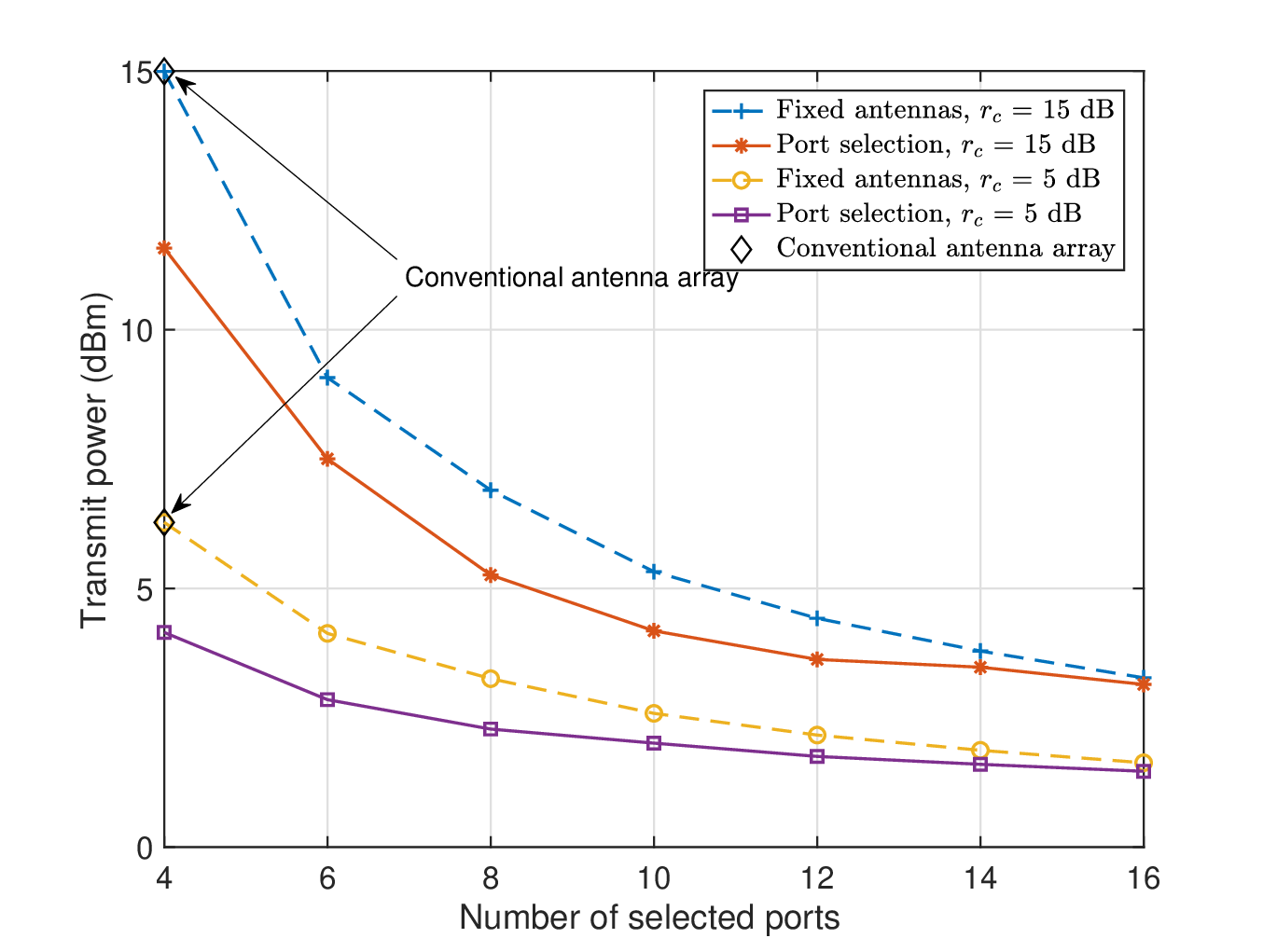}
	\caption{Transmit power versus different number of activated ports $N$, compared with the baseline methods. The sensing SNR threshold is set to $r_s$ = 5 dB. }
	\label{fig:ports}
\end{figure}

In Fig~\ref{fig:ports}, we numerically evaluate the effectiveness of the proposed port selection algorithm for FAS-ISAC. Firstly, compared with the conventional antenna array where only 4 ports can be implemented within the limited space, FAS achieves 33\% power reduction even with 4 selected ports when $r_c = 5$ dB. When the number of activated ports of FAS increases,  it can be noted that the transmit power significantly decreases. This is because the fact that utilizing a higher number of ports provides additional DoF (i.e., spatial diversity gain), thereby achieving a higher integration gain at a certain port. To further show the performance gain of the port selection, we compare the proposed port selection method with the fixed antennas. The fixed antenna method also minimizes the transmit power with the same constraints as the proposed port selection method, but the antennas are fixed and uniformly distributed. 
%Specifically, the beamforming scheme is based on SCA as described in Sec. III. 
In Fig.~\ref{fig:ports}, we can observe a noticeable drop in the transmit power of the proposed port selection method, as the spatial diversity gain is further enhanced due to the flexibility of the port position, leading to a decrease in the required transmit power to meet the S\&C requirements.

\begin{figure}
	\centering
	\includegraphics[width=0.95\linewidth]{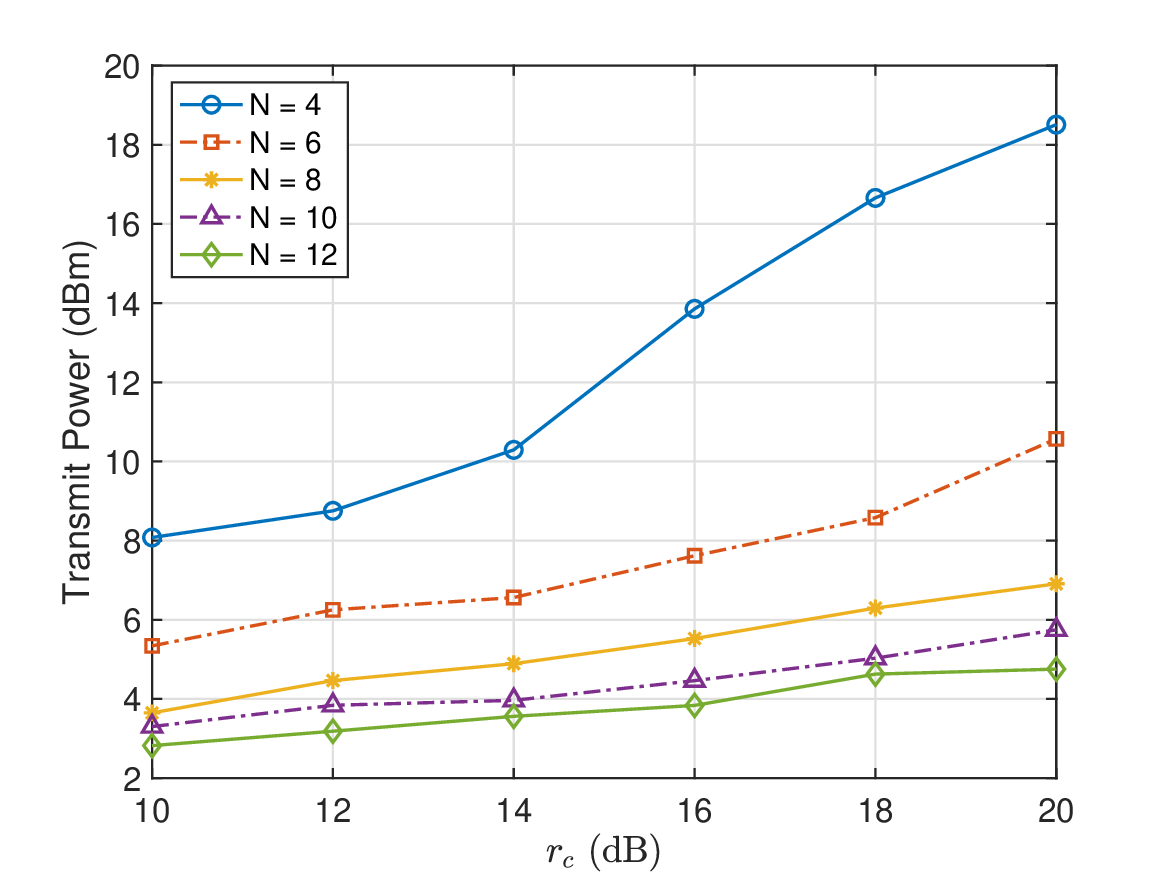}
	\caption{Transmit power versus different communication SNR thresholds $r_{c}$, under different number of selected ports $N$. The sensing SNR threshold is set to $r_s$ = 5 dB.}
	\label{fig:snrc}
\end{figure}
Fig.~\ref{fig:snrc} demonstrates the transmit power versus different communication SNR requirements. As expected, the transmit power increases with the increasing communication SNR requirements in all cases of different available ports. It is worth noting that when the antenna number increases from 4 to 10, only less than half of the transmit power is required to meet the preset SNR thresholds. 
This indicates the potential of the fluid antenna in power saving as it is capable of providing more available ports within a limited space.

\begin{figure}
	\centering
	\includegraphics[width=0.95\linewidth]{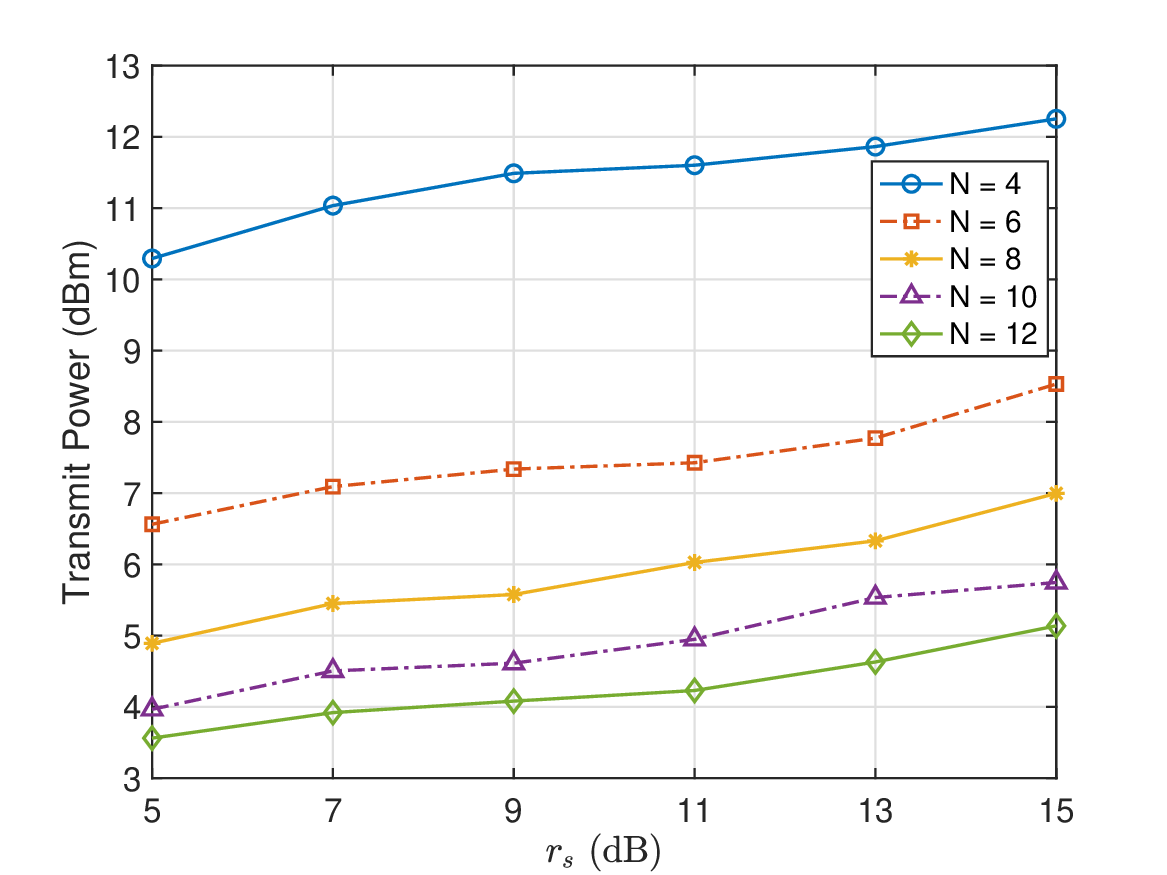}
	\caption{Transmit power versus different sensing SNR thresholds $r_{s}$, under different number of selected ports $N$. The communication SNR threshold is set to $r_{c}$ = 14 dB.}
	\label{fig:snrr}
\end{figure}

We further demonstrate the transmit power versus different sensing SNR thresholds in Fig.~\ref{fig:snrr}. Similar to the trend in Fig.~\ref{fig:snrc}, the transmit power is significantly decreased when more ports are activated. 
Although higher transmit power is required to meet a higher sensing SNR threshold, the increase in power for a larger number of activated ports, e.g., $N$ = 12, is considerably smaller than that for fewer activated ports, e.g., $N$ = 4. This observation suggests the effectiveness of FAS in enhancing sensing capabilities by providing an additional spatial DoF, as the FAS supports a greater number of antennas than conventional antennas within a given array size.

\begin{figure}
	\centering
	\includegraphics[width=0.95\linewidth]{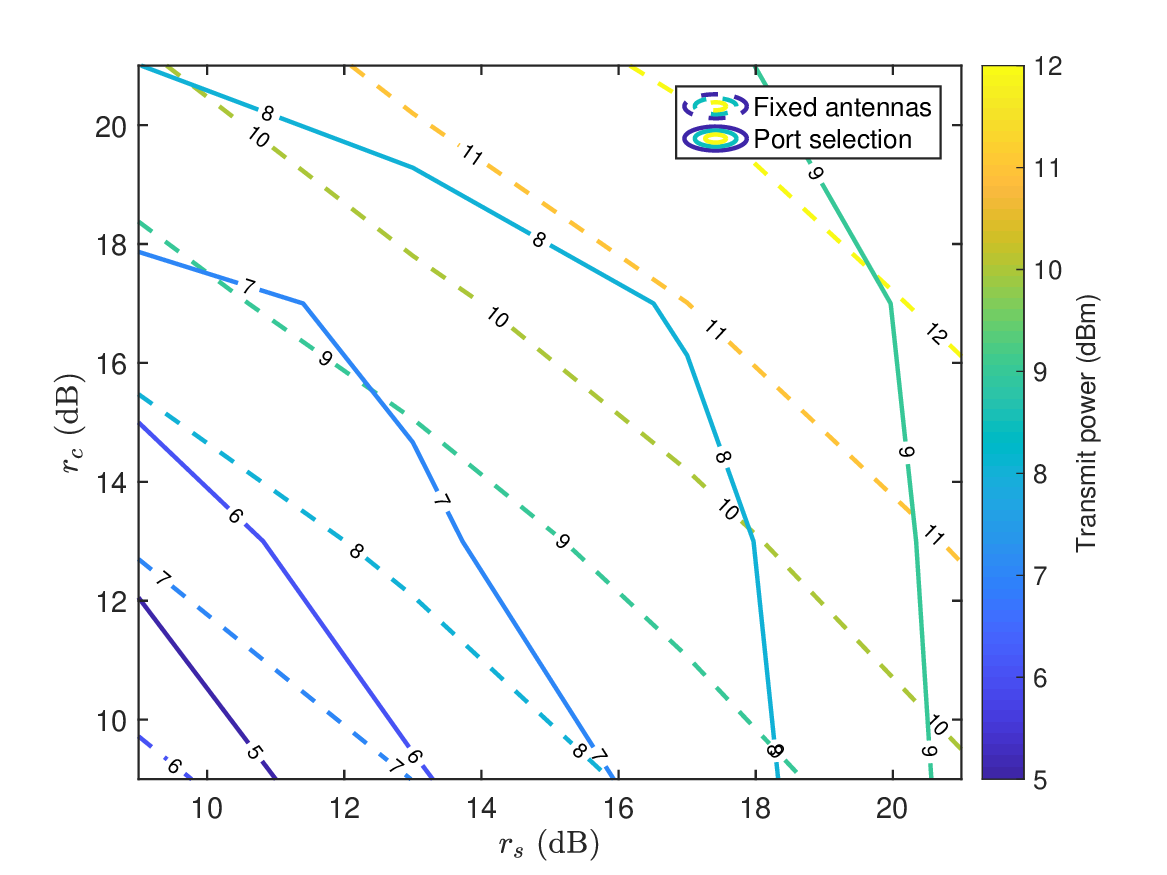}
	\caption{Sensing-communication trade-off in the considered FAS-ISAC system with $N = 8$. The achieved minimum transmit power is labeled on the contour curve.}
	\label{fig:tradeoff}
\end{figure}

To investigate the trade-off between S\&C, we adjust the threshold of communication SNR and sensing SNR, i.e., $r_c$ and $r_s$, and solve a sequence of problems as outlined in \eqref{nonconvexconstraintJoint}. Subsequently, we present the contour curve of the achieved minimum transmit power under varying values of $r_c$ and $r_s$, as shown in Fig.~\ref{fig:tradeoff}. This figure also reveals the S\&C tradeoff of the proposed FAS-ISAC system. Under the same S\&C constraints, the transmit power of FAS-ISAC is much lower than that of the baseline method with fixed antennas, demonstrating a better trade-off between S\&C. These results indicate a significant potential for shifting the ISAC tradeoff and increasing the performance bound via the implementation of FAS.

\balance 
%\textbf{ The inherent conflict between $r_c$ and $r_s$ also appeals a strategic approach for the designers to strike an effective balance along the frontier. }
%\textbf{We further demonstrate the transmit power versus different sensing SNR thresholds in Fig.~\ref{fig:snrr}. Similar to the trend in Fig.~\ref{fig:snrc}, the transmit power is significantly decreased when more ports are activated. Although higher transmit power is required to meet a higher sensing SNR threshold, the increase in power for a larger number of activated ports, e.g., $N$ = 12, is considerably smaller than that for fewer activated ports, e.g., $N$ = 4. This observation suggests the efficacy of FAS in facilitating sensing functionalities. This observation suggests the efficacy of FAS in facilitating sensing functionalities. This observation suggests the efficacy of FAS in facilitating sensing functionalities.}

\section{Conclusion}
In this letter, we studied the model of FAS-enabled ISAC, based on which we proposed a joint port selection and beamforming design.
In particular, we considered an ISAC base station with a FAS, simultaneously performing multicast communication and target sensing. 
Our objective was to design a sparse beamforming vector of minimum power that meets both the communication SNR for users and the sensing SNR for target sensing.
Finally, its superior performance was confirmed by the simulation results, showing that our proposed method can achieve much better performance than both the conventional antenna and uniformly distributed antennas. The potential of shifting the ISAC trade-off with FAS was verified, demonstrating the efficiency of FAS in fostering S\&C performance.
%	\section*{Appendix A}
%
%}

\bibliographystyle{IEEEtran}
\bibliography{isac_fluid}

\vspace{12pt}

\newpage

\vfill

\end{document}